\documentclass[aps, prb, reprint, showpacs, twocolumn, 10pt]{revtex4-1}

\usepackage{graphicx, amsmath, txfonts, bm, braket, ulem, bbold, color}

\begin{document}

\bibliographystyle{apsrev4-1}

\title{Spin-orbit coupling in quasi-one-dimensional Wigner crystals}

\author{Viktoriia Kornich}
\affiliation{Physics and Materials Science Research Unit, University of Luxembourg, L-1511 Luxembourg}
\author{Christopher J. Pedder}
\affiliation{Physics and Materials Science Research Unit, University of Luxembourg, L-1511 Luxembourg}
\author{Thomas L. Schmidt}
\affiliation{Physics and Materials Science Research Unit, University of Luxembourg, L-1511 Luxembourg}

\date{\today}

\begin{abstract}
We study the effect of Rashba spin-orbit coupling (SOC) on the charge and spin degrees of freedom of a quasi-one-dimensional (quasi-1D) Wigner crystal. As electrons in a quasi-1D Wigner crystal can move in the transverse direction, SOC cannot be gauged away in contrast to the pure 1D case. We show that for weak SOC, a partial gap in the spectrum opens at certain ratios between density of electrons and the inverse Rashba length. We present how the low-energy branch of charge degrees of freedom deviates due to SOC from its usual linear dependence at small wave vectors. In the case of strong SOC, we show that spin sector of a Wigner crystal cannot be described by an isotropic antiferromagnetic Heisenberg Hamiltonian any more, and that instead the ground state of neighboring electrons is mostly a triplet state. We present a new spin sector Hamiltonian and discuss the spectrum of Wigner crystal in this limit.
\end{abstract}


\maketitle

\let\oldvec\vec
\renewcommand{\vec}[1]{\ensuremath{\boldsymbol{#1}}}

\section{Introduction}
\label{sec:Introduction}
Low-dimensional systems are of great interest in condensed matter physics because of their broad range of technological applications.\cite{babinec:nnano10, vdovin:prl16} Systems such as quantum dots, nanowires, and two-dimensional electron gases are usually formed using metallic gates and band engineering.\cite{drexler:prl94, yacoby:prl96, laird:prb10, kawakami:nnano14, scheller:prl14} Both of these factors induce a structural asymmetry in the system and, as a consequence, generate spin-orbit coupling (SOC).\cite{winkler:book} The effect of SOC is crucial for many proposed technological applications, for instance in the field of spintronics. For example, SOC can be used as a means to control the spin state of an electron in a quantum dot\cite{nowack:science07} or it can lead to the formation of Majorana fermions in nanowire-superconductor hybrid structures.\cite{oreg:prl10, lutchyn:prl10} This latter property has triggered a lot of experimental research into one-dimensional (1D) systems with SOC.\cite{mourik:science12, deng:nlet12, das:nphys12, rokhinson:nphys12, albrecht:nature16}

It was shown theoretically that when the electron density in a nanowire is very low, it becomes energetically favourable for electrons to arrange in a quasi-long range ordered state: a Wigner crystal.\cite{wigner:pr34, meyer:jpcm09} The charge and spin degrees of freedom in such a system decouple from one another, and so display the same spin-charge separation seen in one-dimensional metallic systems at higher densities where the Luttinger liquid model applies.\cite{haldane:jpc81} There are experimental indications of the presence of Wigner crystals in quantum wires and carbon nanotubes.\cite{hew:prl09, yamamoto:prb12, yamamoto:jpsj15, deshpande:natphys08} A strict one-dimensional arrangement is favored by strong confinement in the transversal direction. In a strictly one-dimensional system, Rashba SOC, which couples the spin and charge modes, can be gauged away using a unitary transformation. As a result, in such a 1D system SOC has no effect on the energy spectrum in the absence of a magnetic field, and spin-charge separation is restored.

If, on the other hand, the confining potential is made shallower, a transition to a quasi-1D zigzag form can take place.\cite{chaplik:pzetf80, chaplik:jetpl80, hasse:ap90, piacente:prb04, meyer:jpcm09, silvestrov:prb14} Moreover, the spectrum of quasi-1D systems is strongly affected by SOC because the latter leads to avoided crossings between neighboring subbands.\cite{moroz:prb99, moroz:prl00, moroz:prb00} For these reasons, Rashba SOC can have a strong effect on the electronic properties of quasi-1D systems even without an applied magnetic field. It is thus important to study this model as it can provide insights into the behavior of conductance and other characteristics of nanowires with Rashba SOC.\cite{schmidt:prb13,pedder:prb16}

The spectrum of a quantum wire with Rashba SOC and applied magnetic field in the limit of strong electron-electron interaction was considered in Ref.~[\onlinecite{schmidt:arxiv16}]. There it was shown that a partial ``helical'' gap in the spectrum can open at certain values of the electron density and this dependence differs from the regimes of non-interacting or weakly interacting electrons. In Ref.~[\onlinecite{pedder:prb16}] it was shown that a helical gap can open in a quasi-1D wire even without an external magnetic field, and due only to electron-electron interactions and Rashba SOC. This result was discovered in the framework of Luttinger liquid theory, which does not describe electronic systems at low densities well.\cite{fiete:rmp07} To accurately describe such a system in the low density limit, we use the more appropriate model of a Wigner crystal.\cite{meyer:jpcm09}

In this article we study strongly-interacting electrons at low densities, which form a quasi-1D Wigner crystal, and investigate the effect of Rashba SOC on the spectrum of such a state, and on its spin and charge degrees of freedom. We consider two cases of particular interest. Firstly, we examine the limit of weak SOC, when it can be treated as a perturbation to the existing description of charge and spin sectors of a Wigner crystal.\cite{meyer:jpcm09} We then investigate the regime of strong SOC, where the effect of Rashba SOC is stronger than that of exchange between neighboring spins. In this case, the spin sector is affected so much that we have to derive a new Hamiltonian for it.

In the regime of weak SOC we first average out the charge degrees of freedom and show that due to SOC the resulting spin Hamiltonian is of XXZ type. This brings the Wigner crystal into the gapped Ising antiferromagnetic regime instead of the gapless isotropic antiferromagnet found without SOC present. We also investigate the charge degrees of freedom of a quasi-1D Wigner crystal in the presence of spins which are classically frozen in the ground state of the unperturbed spin sector Hamiltonian. As a consequence of the zigzag structure, we find four oscillator branches and as we are interested in the low-energy physics, we study the spectrum of the lowest branch. We show that due to SOC the small-$k$ spectrum of this branch deviates markedly from the linear behavior observed in the absence of SOC.

For the case of strong SOC we study how the spin sector Hamiltonian changes. We follow the previously-used procedure for calculating the exchange interaction between spins in a Wigner crystal,\cite{matveev:prb04} and consider a double well potential with two electrons in it. We present the new spin interaction Hamiltonian in the presence of SOC, and show that for strong SOC compared to the tunnel coupling between wells, the lowest energy state is approximately a triplet, in contrast to the singlet ground state expected for an isotropic exchange interaction. For strong SOC the usual description of the spin sector of the Wigner crystal by means of an isotropic Heisenberg Hamiltonian does not apply.

The paper is organized as follows. In Sec.~\ref{sec:TheModel} we present the Hamiltonian of our model. In Sec.~\ref{sec:WeakSOC} we study the spectrum of a Wigner crystal with weak SOC and consider spin and charge degrees of freedom in more detail in two respective subsections. We discuss the case of strong SOC in Sec.~\ref{sec:StrongSOC} and derive a new Hamiltonian for the spin sector. Our conclusions follow in Sec.~\ref{sec:Conclusions}. Details of calculation and additional information are provided in the Appendices.

\section{model}
\label{sec:TheModel}

\begin{figure}[tb]
\begin{center}
\includegraphics[width=\linewidth]{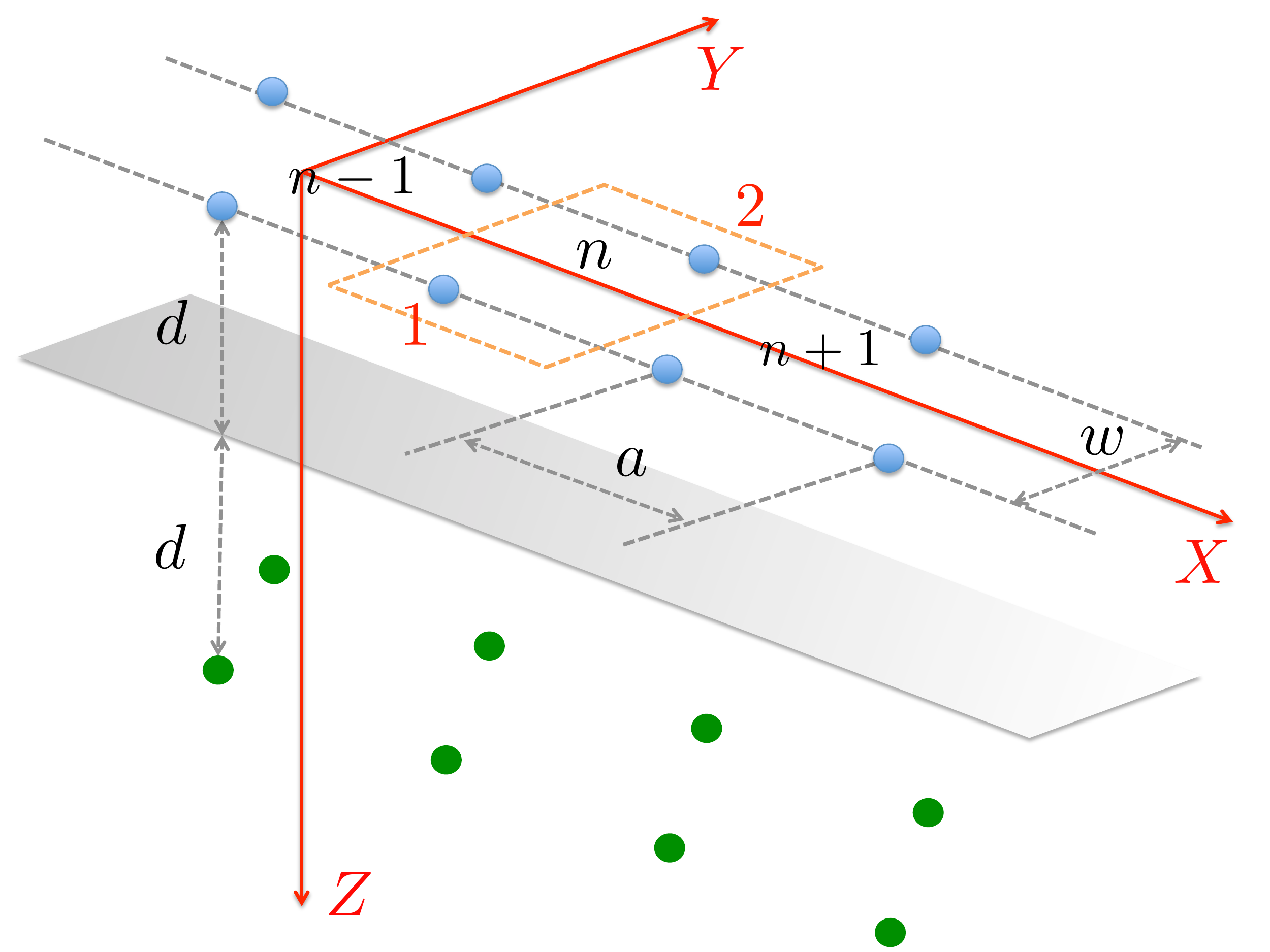}
\caption{(Color online) The arrangement of electrons which are strongly confined in the $Z$-direction, but more weakly confined in the $Y$-direction by a harmonic potential, leading to the zigzag form of the Wigner crystal. The unit cell of the zigzag state is shown in orange, and contains two distinct lattice sites labelled $1$ and $2$. We model the screening of the long-range part of the Coulomb interaction between electrons seen in real experimental systems by means of a metallic gate at a distance $d$ below the confined electrons (blue circles). The presence of this gate results in image charges (green circles) which cause the long-range part of the Coulomb interaction to decay as $1/|X|^3$, as expected for a dipole potential.}
\label{fig:plot_gate_Wigner_crystal}
\end{center}
\end{figure}

We consider electrons confined in the $Y$ and $Z$ directions by an external potential, and which therefore form a one-dimensional structure along the $X$ direction. It is known that if the potential energy due to the Coulomb repulsion between electrons is larger than their kinetic energy, which is typically the case at low densities, it is energetically favorable for electrons to form a 1D lattice with quasi-long-range order, a Wigner crystal.\cite{wigner:pr34} In addition, if the confining potential in the transversal direction is relatively shallow, or the density of electrons is increased, the Wigner crystal can take on a quasi-1D form, a zigzag pattern\cite{chaplik:pzetf80, chaplik:jetpl80, hasse:ap90, piacente:prb04, meyer:jpcm09} as shown in Fig.~\ref{fig:plot_gate_Wigner_crystal}. This phase was reviewed in detail in Ref.~[\onlinecite{meyer:jpcm09}]. Further increasing the density can result in many-row zigzag structures, before the crystal melts due to the enhanced quantum fluctuations at higher densities. The relation between confinement length and dimensions of Wigner crystal is discussed in Appendix \ref{app:HarmonicPotentialApproximation}.

To form 1D systems metallic gates are often used. As a rule, they induce structural asymmetry in the system, which in turn induces Rashba spin-orbit coupling.\cite{winkler:book} We include this in the Hamiltonian of the system, which reads
\begin{equation}
H=\sum_n \frac{(p^n_X)^2+(p^n_Y)^2}{2m}+V(X_n,Y_n)+\alpha(p_X^n\sigma_Y^n-p_Y^n\sigma_X^n),
\end{equation}
where $m$ is the effective mass of electron, $X_n$ and $Y_n$ describe the position of the $n$th electron, $p^n_{X(Y)}$ is the $X$($Y$) component of the electron momentum, and $\sigma^n_{X(Y)}$ is the $X$($Y$) component of its spin. Moreover, $\alpha$ denotes the SOC strength, and $V(X_n, Y_n)$ is the Coulomb interaction between the electrons.

Aside from inducing SOC, the metallic gates also screen the long range part of the Coulomb interaction, which decays at large distances not as $1/|X|$ but rather as $1/|X|^3$. To incorporate this effect into our model, we consider a metallic gate at a distance $d$ beneath the Wigner crystal along the $Z$ direction, see Fig.~\ref{fig:plot_gate_Wigner_crystal}. As a result of this screening, to model the charge excitations of the Wigner crystal we need only include nearest-neighbour interactions between electrons. It was shown that in this case the low-energy charge excitations of Wigner crystal can be described in terms of density waves.\cite{meyer:jpcm09, matveev:prl04}

The equilibrium position of the $n$th electron along a zigzag chain with longitudinal spacing $a$ and width $w$ is given by $(a n, (-1)^n w/2)$. Allowing for small fluctuations, we can express the position of the $n$th electron as $(X_n,Y_n)=(an+x_n, (-1)^n w/2+y_n)$, where $x_n$, $y_n$ are the deviations of the electron from its equilibrium position. We expand $V(X_n,Y_n)$ to second order in $(x_n,y_n)$. The condition for equilibrium is that the first order term vanishes, so that the lowest non-trivial term is of second order. In the limit $d\gg a\gg w$ the form of the potential energy is,
\begin{equation}
\label{eq:VSimplified}
\bar{V}(x_n,y_n)=\frac{m\Theta^2}{2}(x_n-x_{n+1})^2-\frac{m\Omega_1^2}{2}(y_n-y_{n+1})^2+\frac{m\Omega_2^2}{2}y_n^2.
\end{equation}
For the details of the derivation of Eq.~(\ref{eq:VSimplified}) see Appendix~\ref{app:HarmonicPotentialApproximation}. The part of the Hamiltonian which describes the charge sector reads
\begin{eqnarray}
\label{eq:Hc}
H_c=\sum_n\frac{(p_x^n)^2+(p_y^n)^2}{2m}+\bar{V}(x_n,y_n).
\end{eqnarray}
where $p^n_{x,y} = p^n_{X,Y}$.

The low-energy excitations of the spin sector of a Wigner crystal are usually described by the Heisenberg Hamiltonian\cite{meyer:jpcm09, matveev:prl04}
\begin{eqnarray}
\label{eq:HeisenbergHamiltonian}
H_s=\sum_n J{\bm\sigma}_n\cdot{\bm\sigma}_{n+1},
\end{eqnarray}
where $\bm\sigma_n$ denotes the spin of the $n$th electron. Due to the strong Coulomb repulsion between nearest neighbors, the energy barrier for exchange between neighboring electrons is high, and $J$ is exponentially suppressed in the separation between electrons. For 1D Wigner crystals $J>0$, i.e., the energetically favored spin state is one with antiferromagnetic order.\cite{matveev:prl04, matveev:prb04, klironomos:prb05, fogler:prb05} For a zigzag chain with $a\gg w$, the Heisenberg Hamiltonian Eq.~(\ref{eq:HeisenbergHamiltonian}) with $J>0$ remains a good model.\cite{meyer:jpcm09} However, SOC explicitly breaks spin-charge separation, and also modifies the spectrum of quasi-1D systems.\cite{moroz:prb99} The question therefore arises whether we can still consider Eq.~(\ref{eq:HeisenbergHamiltonian}) as describing spin degrees of freedom even when we include SOC in the system.

The exchange interaction between spins in a Wigner crystal is usually derived by considering the exchange of two electrons placed in the Coulomb potential of all the other electrons, and in an external confinement potential.\cite{matveev:prb04} The exchange interaction between two localized electrons in a material with SOC was considered in Ref.~[\onlinecite{kavokin:prb01}]. For weak SOC compared to the exchange interaction $J$, to leading order the spin Hamiltonian retains the form of a Heisenberg Hamiltonian. Consequently, we consider SOC as a perturbation to $H_s+H_c$ when $m\alpha^2\ll J,\Omega_1,\Omega_2,\Theta$, and study the effect of weak SOC on the spectrum of the Wigner crystal. Furthermore, we investigate the case of strong SOC, when the spin excitations of the Wigner crystal cannot be described using $H_s$, and derive a new spin sector Hamiltonian.

\section{Weak Spin-orbit coupling}
\label{sec:WeakSOC}
\subsection{Averaging out the charge degrees of freedom}
\label{subsec:AveragingOutCharge}

In this section we assume that $m\alpha^2\ll J,\Omega_1,\Omega_2,\Theta$, and consider SOC as a perturbation to $H_s+H_c$. To simplify our analysis, we perform a unitary transformation on the Hamiltonian $H \rightarrow U_\sigma^\dagger HU_\sigma$, with $U_\sigma=\prod_n e^{-im\alpha\sigma_Y^nX_n}$. Going over to the Wigner crystal representation, we obtain again the Hamiltonian $H_c+H_s$ as well as an SOC correction coupling spin and charge modes,
\begin{eqnarray}
\label{eq:HSOC}
\nonumber
H_{SOC}=-\alpha p_y^n\Big[\sigma_x^n\cos{2m\alpha(an+x_n)}\\ +\sigma_z^n\sin{2m\alpha (an+x_n)}\Big].
\end{eqnarray}
To study the effect of this term on the spectrum of the Wigner crystal, we consider the regime when $2m\alpha a =\pi $, i.e., a separation between electrons $a$ which is commensurate with the spin-orbit length $\ell_{SO} = (2m \alpha)^{-1}$. This is the same condition for the opening of a helical gap in the spectrum given in Refs.~[\onlinecite{schmidt:arxiv16}] and [\onlinecite{pedder:prb16}]. Since the fluctuations about the equilibrium positions are small, we approximate $\cos(2m\alpha x_n)\simeq 1$, and neglect altogether the term containing $\sin(2m\alpha x_n)$. As a result, our perturbation contains only the $\sigma_x$ component of spin.

To develop a better understanding of the effect of SOC, we average out the charge degrees of freedom and so derive a new effective spin Hamiltonian. We define the partition function $\mathcal{Z}$, which is expressed in terms of the action of the system in imaginary time as
\begin{eqnarray}
\mathcal{Z}=\int \mathcal{D} g\ \mbox{Exp}[-(S_s+S_c+S_{SOC})],
\end{eqnarray}
where the integral measure $g$ includes both spin and charge degrees of freedom, and $S_s$ and $S_c$ describe the dynamics of the spin and charge sectors respectively, in the unperturbed system. $S_{SOC}$ is treated as a perturbation, defined as $S_{SOC}=\int_0^\beta d\tau H_{SOC}$, where $\beta = 1/T$ is the inverse temperature. We expand $\exp(-S_{SOC})$ to second order in $\alpha$, integrate out the charge degrees of freedom and then re-exponentiate the result again. As $S_c$ is quadratic in $y_n$ and $S_{SOC}$ is linear, the average of the linear term in $S_{SOC}$ vanishes, $\langle S_{SOC}\rangle=0$, where $\langle...\rangle=\int\mathcal{D}g_c...\mbox{Exp}[-S_c]$ and $g_c$ denotes charge degrees of freedom. The partition function after integrating out charge modes reads
\begin{eqnarray}
\label{eq:ZAveraged}
\mathcal{Z}\simeq \int \mathcal{D} g_s\ \mbox{Exp}[-S_s+\frac{1}{2}\langle S_{SOC}S_{SOC}\rangle],
\end{eqnarray}
where $g_s$ denotes spin degrees of freedom.

To calculate $\langle S_{SOC}S_{SOC}\rangle$ we transfer to Fourier representation using the transformation \begin{equation}
x_n=\frac{1}{\beta N}\sum_{k,\omega}x_{k,\omega}e^{-ikan+i\omega\tau}
\end{equation}
for all operators. Here $N=L/a$ is the total number of electrons in the Wigner crystal, which has length $L$, and $\omega=2\pi l/\beta$ is a bosonic Matsubara frequency with $l\in \mathbb{Z}$. After averaging out the charge degrees of freedom, we get
\begin{equation}
\langle S_{SOC}S_{SOC}\rangle=-\frac{\alpha^2m}{\beta N}\sum_{k,\omega}\frac{\omega^2\sigma_x^{k,\omega}\sigma_x^{-k,-\omega}}{\omega^2-4\Omega_1^2\cos^2{(ka/2)}+\Omega_2^2}.
\end{equation}
We return to the temporal representation of the spin operators as follows. We assume that $J\ll \Theta, \Omega_1,\Omega_2$, so that the charge degrees of freedom evolve on much faster time scales than the spin degrees of freedom. This assumption is justified by the fact that $J$ decays exponentially with the distance between electrons,\cite{matveev:prb04} whereas $\Omega_1$ and $\Theta$ decay as power laws as a function of distance with exponent $-3/2$ (see Appendix \ref{app:HarmonicPotentialApproximation}), and $\Omega_2$ is required to be larger than $\Omega_1$ to ensure the stability of the Wigner crystal. We then integrate the prefactor multiplying $\sigma_x^{k,\tau}\sigma_x^{-k,\tau'}$ over $(\tau-\tau')$, and approximate the result by an instantaneous interaction between the spins. The non-trivial term becomes
\begin{eqnarray}
\label{eq:CorrectionKTau}
\nonumber
\langle S_{SOC}S_{SOC}\rangle=-\frac{\alpha^2m}{N}\int_0^\beta d\tau\sum_{k,n,n'} e^{-\beta\sqrt{\Omega_2^2-4\Omega_1^2\cos^2{[ka/2]}}} \\ \times e^{ika(n-n')}\sigma_x^{n,\tau}\sigma_x^{n',\tau}.\ \ \ \
\end{eqnarray}
From this expression we see that if the temperature is zero ($\beta\rightarrow \infty$) the average vanishes: at zero temperature the spin-orbit coupling produces no nearest-neighbor spin exchange up to second order in $\alpha$. A second feature of this formula is that if $\Omega_2\rightarrow\infty$ so that oscillations in $Y$ direction are forbidden, the average also becomes zero. This means that in a strictly 1D Wigner crystal this SOC-induced correction is absent, which is compatible with the fact that SOC can be gauged away in the 1D limit.

To express $\langle S_{SOC}S_{SOC}\rangle$ in the spatial representation, we consider the case $\beta[\Omega_2^2-4\Omega_1^2\cos^2{(ka/2)}]^{1/2}\ll 1$, which corresponds to large enough temperatures (small $\beta$) and shallow enough external potential (small $\Omega_2$). We expand the exponential in Eq.~(\ref{eq:CorrectionKTau}) and carry out the summation over $k$. The resulting expression is non-zero only when $n=n'$ and $n-n'=\pm 1$. As the $n=n'$ term corresponds to a constant shift in the action, we ignore it, and consider only the $n-n'=\pm 1$ terms. Then the average is
\begin{equation}
\langle S_{SOC}S_{SOC}\rangle=-\alpha^2m\beta\frac{\Omega_1^2}{\Omega_2}\int_0^\beta d\tau\sum_{n}\sigma_x^{n,\tau}\sigma_x^{n+1,\tau}.
\end{equation}
We put this back into Eq.~(\ref{eq:ZAveraged}) to arrive at an effective spin Hamiltonian which reads
\begin{equation}
\label{eq:tildeHs}
\tilde{H}_s=\sum_n J{\bm \sigma}_n\cdot{\bm \sigma}_{n+1}+\alpha^2m\beta\frac{\Omega_1^2}{2\Omega_2}\sigma_x^{n}\sigma_x^{n+1}.
\end{equation}
This Hamiltonian describes an XXZ-type Heisenberg chain because the prefactor of $\sigma_x^n\sigma_x^{n+1}$ is different from the coefficient multiplying $\sigma_y^n\sigma_y^{n+1}$ and $\sigma_z^n\sigma_z^{n+1}$. In addition, since the coefficient of the $\sigma_x^n\sigma_x^{n+1}$ term is larger than that for the other spin directions, the Hamiltonian (\ref{eq:tildeHs}) is in the Ising antiferromagnet regime, which has a gapped spectrum.\cite{giamarchi:book}

One can show that this gap is present not only at electron density $2m\alpha a=\pi$, but at all commensurate densities defined by $2m\alpha a=(2l+1)\pi$. In contrast, if $2m\alpha a=2\pi l$, the spectrum is gapless because the correction has the opposite sign, so our system is in the XY phase.\cite{giamarchi:book} For the details of the calculation for arbitrary $2m\alpha a$ see Appendix \ref{app:ArbitraryDensity}.

\subsection{Spectrum of charge degrees of freedom}
\label{subsec:SpectrumCharge}

In the previous section, we integrated out the charge degrees of freedom to second order in $\alpha$. At $T=0$, the spin degrees of freedom were unaffected by SOC. Since SOC explicitly couples spin and charge degrees of freedom, this naturally leads us to investigate if the effect of SOC can instead be seen in the charge degrees of freedom at $T=0$. Therefore, in this section we shall assume the spins to be frozen in the classical, N\'eel ground state of the isotropic Heisenberg model, whereas the charge degrees of freedom are still able to fluctuate.

We begin from the description of the Wigner crystal in zigzag form, where we have a unit cell that contains two electrons as shown in Fig.~\ref{fig:plot_gate_Wigner_crystal}. First of all we simplify the form of SOC by performing a Schrieffer-Wolff transformation up to second order in $\alpha$. We use the Schrieffer-Wolff transformation $e^{iU_{SW}}He^{-iU_{SW}}$, where the hermitian operator $U_{SW}$ reads\cite{khaetskii:prb00, aleiner:prl01}
\begin{equation}
U_{SW}=m\alpha \sum_{n,\gamma} \left[ \sigma_X^{n,\gamma} Y_n^\gamma-\sigma_Y^{n,\gamma} X_n^\gamma \right].
\end{equation}
Here $\gamma$ denotes the type of the electron in the unit cell: first ($\gamma=1$) or second ($\gamma=2$), as shown in Fig.~\ref{fig:plot_gate_Wigner_crystal}. To lowest order in $\alpha$ our SOC Hamiltonian becomes
\begin{equation}
H_{SOC}^{SW}=-m\alpha^2\sum_{n,\gamma}\sigma_Z^{n,\gamma}\left[Y_{n,\gamma}p_X^{n,\gamma}-X_{n,\gamma}p_Y^{n,\gamma}\right].
\end{equation}
To this order in $\alpha$, $\sigma_Z^{n,\gamma}$ is conserved. Therefore, we make the ansatz that the spins are frozen in the antiferromagnetic ordering corresponding to the classical lowest energy state of Eq.~(\ref{eq:HeisenbergHamiltonian}), and so take $\sigma_Z^{n,\gamma}=(-1)^{\gamma-1}$.

\begin{figure}[tb]
\begin{center}
\includegraphics[width=\linewidth]{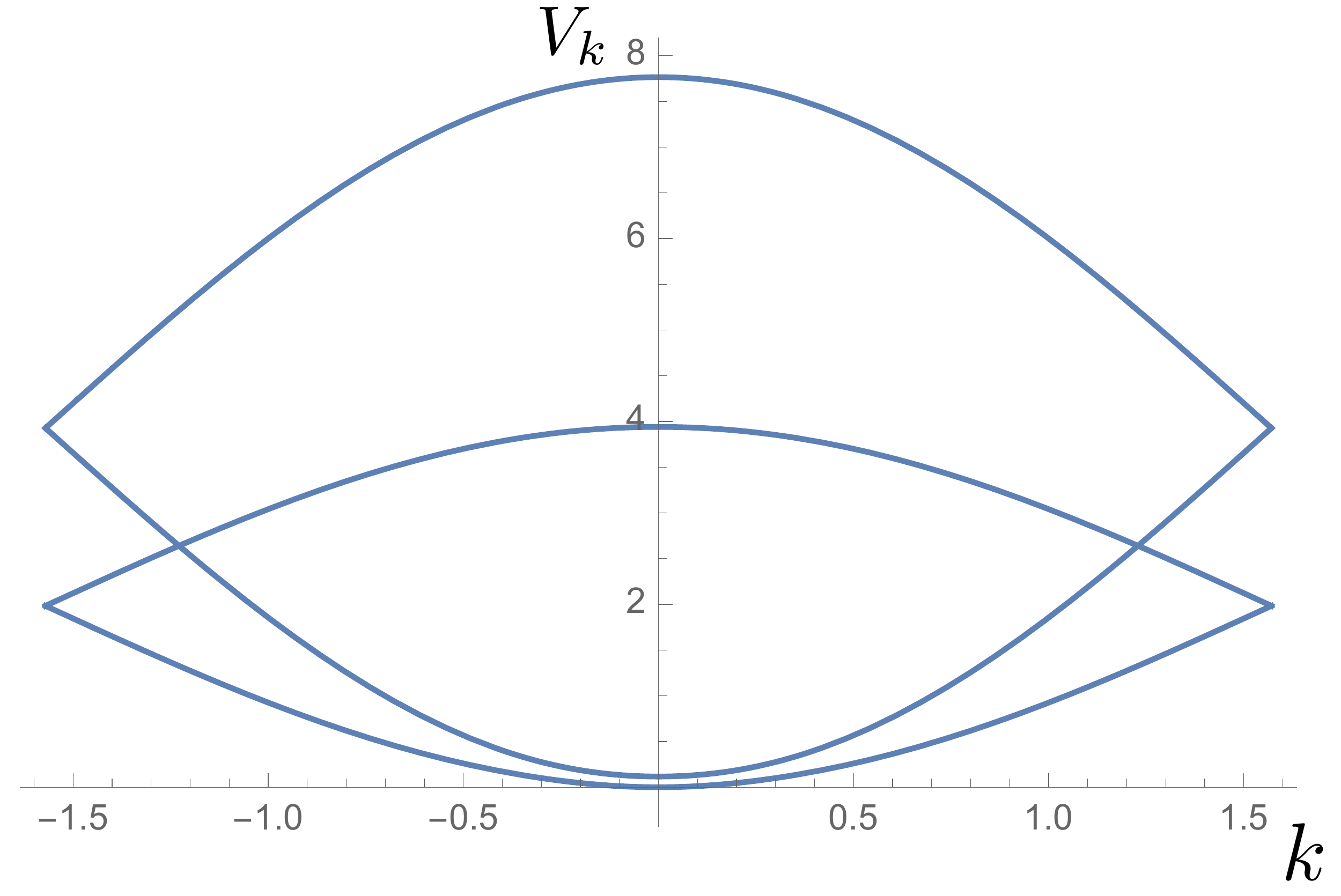}
\caption{(Color online) The eigenvalues of the matrix $V_k$ as a function of $k$ for the four charge-sector normal modes in the zigzag Wigner crystal without SOC present. Whilst the exact parameters for this plot are given in the text, the qualitative behavior of these eigenvalues does not depend sensitively on their values. Focussing on the lowest mode, we see that this eigenvalue, which corresponds to the frequency squared of the lowest in energy oscillator is approximately quadratic for small $k$, leading to a linear dispersion relation for this mode.}
\label{fig:diagonalization}
\end{center}
\end{figure}

In the part of the Hamiltonian describing the low-energy charge excitations, we now take the zigzag structure fully into account. The resulting Hamiltonian $H_c^z$ has a similar form to $H_c$ from Eq.~(\ref{eq:Hc}), however it also contains a summation over $\gamma$, and $V(x_n,y_n)$ from Eq.~(\ref{eq:VExpanded}) instead of $\bar{V}(x_n,y_n)$ from Eq.~(\ref{eq:VSimplified}):
\begin{equation}
H_c^z=\sum_n\left[ \left(\sum_\gamma\frac{(p_x^{n,\gamma})^2+(p_y^{n,\gamma})^2}{2m} \right) +V(x_n,y_n)\right].
\end{equation}
To find the spectrum of $H_c^z+H_{SOC}^{SW}$, we define a new Fourier representation which respects the periodicity of the zigzag structure in the following way
\begin{eqnarray}
x_{n,\gamma}&=&\sqrt{\frac{2}{N}}\sum_kx_{k,\gamma} e^{-2ikan},\\
p_x^{n,\gamma}&=&\sqrt{\frac{2}{N}}\sum_kp_x^{k,\gamma}e^{2ikan}.
\end{eqnarray}
We express $V(x_n,y_n)$ in this Fourier representation, and then diagonalize it. We denote the result of the diagonalization as $V_k$, and plot it in Fig.~\ref{fig:diagonalization} for the following parameters: $d=10$, $w=0.1$ in the units of $a$, and $D_\Omega\simeq 3.94$ in the units $e^2/(\epsilon a^3)$ derived from Eq.~(\ref{eq:EquilibriumCondition}).

We denote the transformation which implements the diagonalization at $k=0$ as $U_{d1}$. Then in the basis
\begin{equation}
\begin{pmatrix}Q_{a,1}\\ Q_{b,1}\\ Q_{a,2}\\ Q_{b,2}\end{pmatrix}=U_{d1}^T\begin{pmatrix}x_{k,1}\\ y_{k,1}\\ x_{k,2}\\ y_{k,2}\end{pmatrix},
\end{equation}
we find
\begin{eqnarray}
V_{k=0}=\begin{pmatrix}0 & 0 & 0 & 0\\
0 & 4(D_{xx}^{(1)}-D_{xx}^{(2)}) & 0 & 0 \\
0 & 0 & D_\Omega & 0\\
0 & 0 & 0 & 4D_{yy}^{(1)}-4D_{yy}^{(2)}+D_\Omega\end{pmatrix}.\ \ \ \ \ \end{eqnarray}
In the limit $d\gg a\gg w$ the second eigenstate has an eigenvalue which is approximately $4D_{xx}^{(1)}$ and positive. The fourth eigenstate has eigenvalue approximately $D_\Omega+4D_{yy}^{(1)}$, and since $D_{yy}^{(1)}<0$ it can be rather small. For an estimate we took the same parameters as for Fig.~\ref{fig:diagonalization}, and get that the second eigenvalue is $7.765$ and the fourth is $0.117$ in units of $e^2/(\epsilon a^3)$. Near to $k=0$, the lowest energy states are therefore those that live on the $(a,1)$ and the $(b,2)$ branches. In the following we will consider only the subspace formed by these lowest branches, and study whether SOC-induced mixing can significantly change the spectrum of the lowest branch.

In the same way as we defined the $Q$ normal modes for the coordinates, we define a similar basis of momentum normal modes. We then express the momentum and coordinate operators via ladder operators $a_k$ for $(a,1)$ and $b_k$ for $(b,2)$ as is usually done for quantum harmonic oscillators: $Q_{a,1}=(a_{-k}^\dagger+a_k)/\sqrt{2m\omega_a}$ for coordinate and similarly for the momentum. The Hamiltonian $H_c^z$ for $k\sim 0$ becomes
\begin{eqnarray}
H_c^z(k\sim 0)=\sum_k\omega_a a^\dagger_ka_k+\omega_bb_k^\dagger b_k,
\end{eqnarray}
where using the results of the diagonalization, we take $\omega_a\propto k$ and $\omega_b\sim \text{const}$ for small $k$. We express $H_{SOC}^{SW}$ in this basis too, to get
\begin{eqnarray}
\nonumber
H_{SOC}^{SW}(k\sim 0)=i\sum_k\left[A(b_{-k}^\dagger+b_k)(a_k^\dagger-a_{-k})\right.\\ \left.-B(a_{-k}^\dagger+a_k)(b_k^\dagger-b_{-k})\right],
\end{eqnarray}
where we introduced the shorthand $A=(m\alpha^2/2) \sqrt{\omega_a/\omega_b}$ and $B=(m\alpha^2/2) \sqrt{\omega_b/\omega_a}$.

\begin{figure}[tb]
\begin{center}
\includegraphics[width=0.99\linewidth]{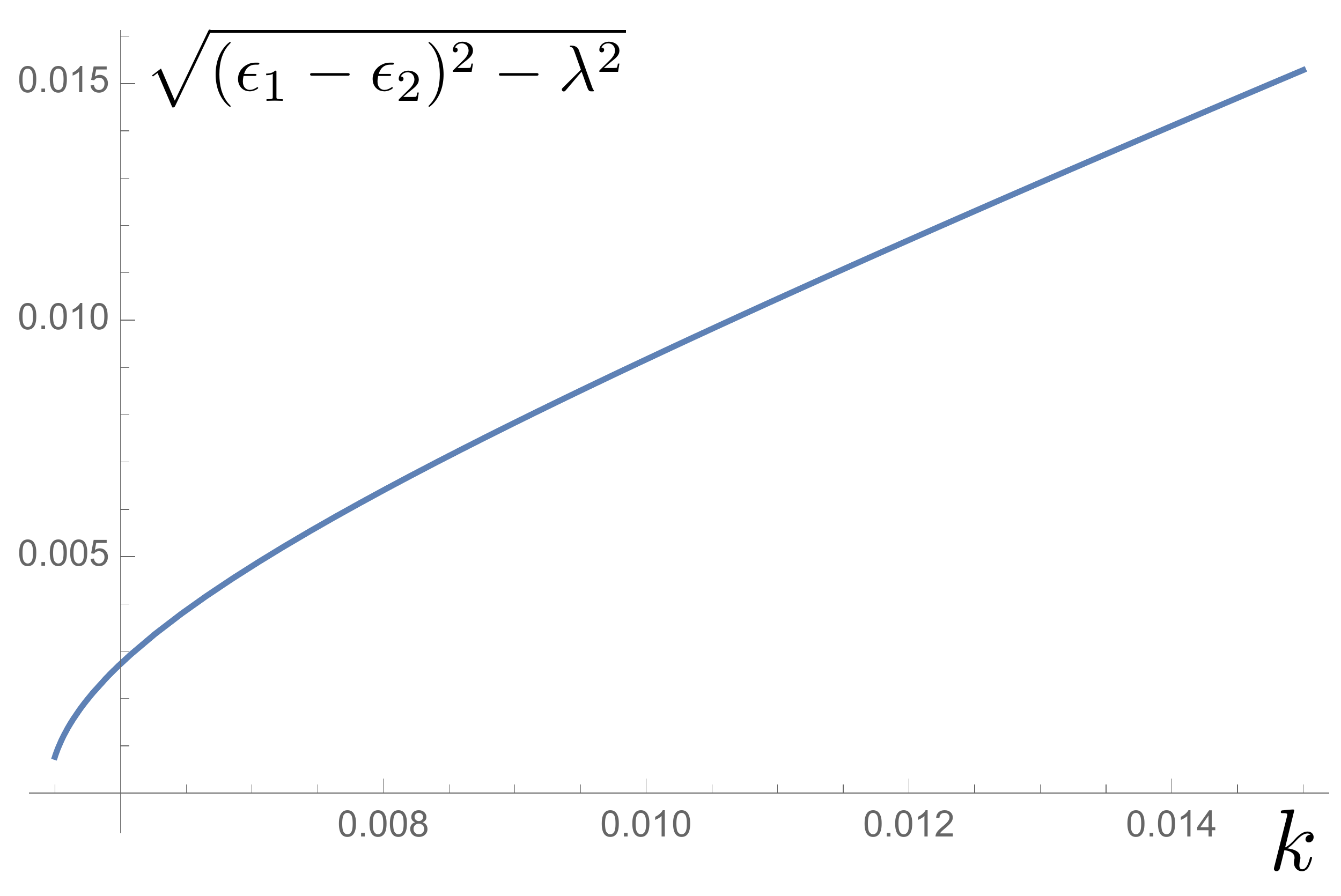}
\caption{The spectrum of the lowest branch of charge degrees of freedom with SOC. Here $\sqrt{(\epsilon_1-\epsilon_2)^2-\lambda^2}$ is in units of $e^2/(\epsilon a)$ and $k$ is in units of $1/a$. The parameters are presented in the text. Here again, as for Fig.~\ref{fig:diagonalization}, the qualitative behavior of $\sqrt{(\epsilon_1-\epsilon_2)^2-\lambda^2}$ is important. We see that $\sqrt{(\epsilon_1-\epsilon_2)^2-\lambda^2}$ noticeably deviates from linear dependence for small $k$.}
\label{fig:spectrum}
\end{center}
\end{figure}

In order to find the spectrum of the lowest branch, we perform several transformations. Firstly, we perform a Bogoliubov transformation with the coefficients $u_1$ and $v_1$ defined as: $u_1=e^{i\phi}\cosh{\theta_1}$, $v_1=e^{i\phi}\sinh{\theta_1}$ with $\phi=\pi/4$, $\tanh{2\theta_1}=2(A-B)/(\omega_a+\omega_b)$. We then block-diagonalize the Hamiltonian, take the lowest-energy block, and perform the second Bogoliubov transformation with the coefficients $u_2$ and $v_2$ defined as: $u_2=\cosh{\theta_2}$, $v_2=\sinh{\theta_2}$, where $\tanh{2\theta_2}=-\lambda/(\epsilon_1-\epsilon_2)$ and
\begin{eqnarray}
\epsilon_1&=&\frac{1}{2}((\omega_a+\omega_b)\cosh{2\theta_1}-2(A-B)\sinh{2\theta_1}),\\
\epsilon_2&=&\frac{1}{2}\sqrt{2(A+B)^2+(\omega_a-\omega_b)^2+2(A+B)^2\cosh{4\theta_1}},\ \ \ \ \ \ \\
\lambda&=&-(A+B)\sinh{2\theta_1}.
\end{eqnarray}
The diagonal term $\sqrt{(\epsilon_1-\epsilon_2)^2-\lambda^2}$, i.e. energy of the lowest branch, reads
\begin{widetext}
\begin{eqnarray}
\nonumber
&&\sqrt{(\epsilon_1-\epsilon_2)^2-\lambda^2}=\\ \label{eq:eigenmodes} &&=\frac{1}{\sqrt{2}}\sqrt{8AB+\omega_a^2+\omega_b^2-\sqrt{(\omega_a+\omega_b)^2-4(A-B)^2}\sqrt{4(A+B)^2+(\omega_a-\omega_b)^2+\frac{16(A+B)^2(A-B)^2}{(\omega_a+\omega_b)^2-4(A-B)^2}}}.\end{eqnarray}
\end{widetext}
The two Bogoliubov transformations impose conditions on our parameters to enforce the reality of the eigenenergies in Eq.~(\ref{eq:eigenmodes}). In particular, these conditions do not allow to consider $k\rightarrow 0$, because in this case $B$ diverges. To better understand the behaviour of $\sqrt{(\epsilon_1-\epsilon_2)^2-\lambda^2}$, we plot it for the same parameters used previously in Fig.~\ref{fig:diagonalization} and $m\alpha^2=0.006$ in units of $e^2/(\epsilon a)$. The allowed interval for $k$ is determined by the conditions imposed by validity of Bogoliubov transformation. In Fig.~\ref{fig:spectrum} we see that the spectrum noticeably deviates from linear dependence for small $k$ in contrast to the linear behavior of $\omega_a$ for small $k$ found previously. We note that the stability of the Wigner crystal is not affected because even though the spectrum changed, the lowest branch remains positive.

\section{Strong spin-orbit coupling}
\label{sec:StrongSOC}
In this section we derive the Hamiltonian for the spin sector of the Wigner crystal which should be used for large SOC, when the Heisenberg Hamiltonian Eq.~(\ref{eq:HeisenbergHamiltonian}) no longer applies. Following Ref.~[\onlinecite{matveev:prb04}] we will study the exchange process between two neighboring electrons of the Wigner crystal. We consider a double well potential which is formed by the Coulomb potential of all the other electrons in Wigner crystal, and by the external confining potential in $Y$ direction. Taking into account that we consider only low-energy excitations, we define the double well potential as\cite{burkard:prb99}
\begin{equation}
V_{dw}(x_n,y_n)=\frac{m\omega_0^2}{2}\left[\frac{1}{a^2}\left(x_n^2-\frac{a^2}{4}\right)^2+y_n^2\right],
\end{equation}
which can be approximated for low energies as two harmonic potentials with frequency $\omega_0$, whose centers are at the distance $a$ from each other. The frequency $\omega_0$ can be determined from the physical, screened Coulomb repulsion between the electrons, $\omega_0=(2D_{xx}^{(1)}/m)^{1/2}$, where $D_{xx}^{(1)}$ is defined in Eq.~(\ref{eq:Dxx1}). Since we took into account all the charge interactions between the electrons, we are left with two spin-$1/2$ fermions placed in the double well potential and subjected to SOC. The Hamiltonian that describes these two fermions reads
\begin{equation}
H_{2f}=\sum_{n=1,2} \left[\frac{(p_x^n)^2+(p_y^n)^2}{2m}+V_{dw}(x_n,y_n)+\alpha (p_x^n\sigma_y^n-p_y^n\sigma_x^n) \right].
\end{equation}
The low-energy subspace for these two fermions includes singlet $|(1,1)S\rangle$ and triplet states $|(1,1)T_0\rangle$, $|(1,1)T_+\rangle$, $|(1,1)T_-\rangle$, where the numbers in parentheses denote the number of electrons in the left and right well. To take into account exchange of fermions, we also include singlets with a doubly occupied left and right well: $|(2,0)S\rangle$ and $|(0,2)S\rangle$. We do not include the corresponding triplet states $|(2,0)T_{0,+,-}\rangle$ and $|(0,2)T_{0,+,-}\rangle$ because to form such a triplet, one fermion must occupy a higher orbital state. Therefore, these states are higher in energy than the singlet states $|(2,0)S\rangle$ and $|(0,2)S\rangle$, and the tunnel coupling between triplets is weaker than for singlets.

Physically, the coexistence of two electrons on the same site of the Wigner crystal is forbidden, and would destroy the double-well potential we consider. As a result, we include $|(2,0)S\rangle$ and $|(0,2)S\rangle$ assuming they have very large energy, and are allowed only as part of a virtual process between $(1,1)$ and $(2,0)$, $(0,2)$ states.

Following Refs.~[\onlinecite{burkard:prb99, stepanenko:prb03, stepanenko:prb12, kornich:prb14}] we define the wave functions of the states described above as:
\begin{eqnarray}
\label{eq:02S}
|(0,2)S\rangle&=&|\Psi_R\rangle|S\rangle,\\
|(2,0)S\rangle&=&|\Psi_L\rangle|S\rangle,\\
|(1,1)S\rangle&=&|\Psi_+\rangle|S\rangle,\\
|(1,1)T_{0,+,-}\rangle&=&|\Psi_-\rangle|T_{0,+,-}\rangle,
\end{eqnarray}
where the spin parts read
\begin{eqnarray}
|S\rangle&=&\frac{|\uparrow\downarrow\rangle-|\downarrow\uparrow\rangle}{\sqrt{2}},\\
|T_0\rangle&=&\frac{|\uparrow\downarrow\rangle+|\downarrow\uparrow\rangle}{\sqrt{2}},\\
|T_+\rangle&=&|\uparrow\uparrow\rangle,\\
\label{eq:Tminus}
|T_-\rangle&=&|\downarrow\downarrow\rangle.
\end{eqnarray}
We choose the quantization axis to be along $Y$. The orbital part of the wave functions is constructed using the ground state wave functions for the harmonic oscillator. The detailed definitions are presented in Appendix~\ref{app:Wavefunctions}.

The Hamiltonian $H_{2f}$ in the basis $|(2,0)S\rangle$, $|(0,2)S\rangle$, $|(1,1)S\rangle$, $|(1,1)T_0\rangle$, $|(1,1)T_-\rangle$, $|(1,1)T_+\rangle$ reads
\begin{eqnarray}
H_{2f}=
\begin{pmatrix} U & 0 & -\sqrt{2}t & i \Xi & 0 & 0\\
0 & U & -\sqrt{2}t & i \Xi & 0 & 0\\
-\sqrt{2}t & -\sqrt{2}t & 0 & 0 & 0 & 0\\
-i \Xi & -i \Xi & 0 & 0 & 0 & 0\\
0 & 0 & 0 & 0 & 0 & 0\\
0 & 0 & 0 & 0 & 0 & 0\end{pmatrix},
\end{eqnarray}
where the tunnel coupling $t$ and the SOC matrix element $\Xi$ are defined as
\begin{align}
t &= -\frac{1}{16ml_c^4}\frac{a^2-12l_c^2}{4 \sinh[a^2/(4l_c^2)]}, \notag \\
\Xi &= \frac{\alpha a}{l_c^2}\frac{1}{\sqrt{2\left(\exp[a^2/(2l_c^2)]-1\right)}}.
\end{align}
and $l_c=1/\sqrt{m\omega_0}$. As previously described, the energy $U$, which plays the role of the on-site repulsion, is very large, and can be roughly estimated as $U\sim e^2/(\epsilon l_c)$, where $\epsilon$ is the dielectric constant, $e$ is the electron charge. The states $|(1,1)T_-\rangle$ and $|(1,1)T_+\rangle$ are not coupled to any other states in this basis, so we omit them, keeping in mind that their energies do not change.

To find the eigenvalues and eigenstates of $H_{2f}$, we diagonalize it to get
\begin{eqnarray}
\label{eq:H2fDiag}
 U_{d2}^{-1} H_{2f}U_{d2}=\begin{pmatrix}
 U & 0 & 0 & 0 \\
 0 & \frac{1}{2}\left(U+\Delta\right) & 0 & 0  \\
 0 & 0 & \frac{1}{2}\left(U-\Delta\right) & 0 \\
 0 & 0 & 0 & 0 \end{pmatrix},
\end{eqnarray}
where $\Delta=\sqrt{16t^2+U^2+8\Xi^2}$, and the transformation matrix $U_{d2}$ reads
\begin{eqnarray}
\nonumber
U_{d2}=\begin{pmatrix}
-1 & \frac{i(U+\Delta)}{4\Xi} & \frac{i(U-\Delta)}{4\Xi} & 0 \\
1 & \frac{i(U+\Delta)}{4\Xi} & \frac{i(U-\Delta)}{4\Xi} & 0 \\
 0 & -\frac{i\sqrt{2}t}{\Xi} & -\frac{i\sqrt{2}t}{\Xi} & \frac{i\Xi}{\sqrt{2}t}\\
 0 & 1 & 1 & 1 &
\end{pmatrix}.\ \ \ \ \
\end{eqnarray}
The first two states have energy close to $U$, so these are mainly a combination of $|(2,0)S\rangle$ and $|(0,2)S\rangle$. The most interesting for us is the lowest energy state, which is a combination of $|(1,1)S\rangle$, $|(1,1)T_0\rangle$, $|(2,0)S\rangle$ and $|(0,2)S\rangle$, and has energy $(U-\Delta)/2$. The contribution from $|(0,2)S\rangle$ and $|(2,0)S\rangle$ scales as $-i\sqrt{2t^2+\Xi^2}/U$, and so is small as $U\gg t,\Xi$. The contribution from $|(1,1)S\rangle$ scales as $-i\sqrt{2}t/\sqrt{2t^2+\Xi^2}$, and the contribution from $|(1,1)T_0\rangle$ scales as $\Xi/\sqrt{2t^2+\Xi^2}$. Whether the lowest energy state is more triplet or singlet therefore depends on the relative strengths of $t$ and $\Xi$. When $\Xi\gg t$, the lowest state is mainly a triplet state. In this case, which corresponds to large SOC, we cannot use the Heisenberg Hamiltonian Eq.~(\ref{eq:HeisenbergHamiltonian}) to describe the spin sector of the Wigner crystal any more.

Comparing our result with that of Ref.~[\onlinecite{kavokin:prb01}], we see that for very large SOC the relative angle by which spins rotate while propagating between lattice sites is $\pi/2$: if we apply the transformation $e^{i \pi \sigma_y /2}$ to $|S\rangle$, we indeed get $|T_0\rangle$.

Using the exact expressions for the wave functions from Eqs.~(\ref{eq:02S})-(\ref{eq:Tminus}) and Eqs.~(\ref{eq:Psipm})-(\ref{eq:PsiLR}), one can express $U_{d2}^{-1} H_{2f}U_{d2}$ in terms of projectors onto orbital states and spin operators. For the case $\Xi \gg t$ we approximate the spin sector Hamiltonian for the low-energy subspace (i.e. without the states with energies around $U$) as
\begin{eqnarray}
\nonumber
\mbox{Tr}_\Phi[U_{d2}^{-1}H_{2f}U_{d2}]=\frac{1}{8}(\Delta-U)\left[-\sigma_x^1\sigma_x^2-\sigma_y^1\sigma_y^2+\sigma_z^1\sigma_z^2\right],
\end{eqnarray}
where $\mbox{Tr}_\Phi[...]$ means trace over orbital states. We see that if we make a transformation $\sigma_x^1\rightarrow -\sigma_x^1$ and $\sigma_y^1\rightarrow -\sigma_y^1$ this Hamiltonian corresponds to the gapless, isotropic antiferromagnetic Heisenberg Hamiltonian. In particular, this means that there is no gap in the spectrum if SOC is very strong.

We note that the effect of SOC on the spin-spin interaction between electrons in the double-well potential was considered in Ref.~[\onlinecite{gangadharaiah:prl08}]. It was shown there that in the absence of overlap between the wave functions of electrons in the different wells, there is an anisotropic spin-spin coupling of the van der Waals type at order $\alpha^4$. In contrast, in our case we assume non-zero tunnel coupling $t$ between the wells, and in this case we found a correction to the ground state energy of order $\alpha^2$ given in Eq.~(\ref{eq:H2fDiag}). We also find that the spin-spin interaction between electrons is an exchange interaction.

\section{Conclusions}
\label{sec:Conclusions}
In this work we studied the effect of SOC on the charge and spin degrees of freedom in a quasi-1D Wigner crystal. We considered two cases: weak SOC which acts as a perturbation to the known description of spin and charge sectors in a Wigner crystal, and strong SOC which changes the spin dynamics profoundly.

As a perturbation, SOC opens a gap to the second order at certain densities of electrons. The gap opens because the correction due to SOC brings the Wigner crystal into the gapped Ising antiferromagnetic regime instead of the gapless, isotropic Heisenberg antiferromagnetic regime found without SOC present. To this order in perturbation theory, a finite temperature is necessary because if the electrons do not move, SOC cannot affect the spins. The potential in the transverse direction should be rather shallow to allow electrons to oscillate in this direction around their equilibrium positions. Otherwise, as in pure 1D systems, SOC can be gauged away and consequently cannot open a gap. The opening of a gap in the spectrum affects many physical properties of the Wigner crystal, e.g. the conductance and the response functions. Therefore our results could be helpful in understanding the behavior of nanowires with spin-orbit coupling.

For the case of weak SOC, we also considered the charge degrees of freedom in more detail. Assuming that the spins are classically frozen into the N\'eel state (in agreement with the antiferromagnetic regime of the unperturbed spin Hamiltonian), we studied the charge degrees of freedom taking into account the zigzag form of quasi-1D Wigner crystal. Out of four oscillator branches the most interesting is the lowest branch because it describes the low-energy excitations. Our results show that the spectrum of this branch noticeably deviates from its linear behaviour for small momenta without SOC.

For the case when SOC is strong, we derived a new spin sector Hamiltonian. We showed that for the case of very strong SOC the lowest energy state is mainly a triplet, so we cannot use the Heisenberg Hamiltonian to describe the spin dynamics any more. We present the evolution of the states of the lowest energy between singlet and triplet character as the relation between SOC and the inter-well tunnel coupling is changed. This analysis provides a spin sector Hamiltonian even at rather large SOC.

\begin{acknowledgments}
We thank T. Meng, D. Stepanenko, C. Psaroudaki, A. Rod, and M. Sadhukhan for helpful discussions. The authors acknowledge support by the National Research Fund Luxembourg
(ATTRACT 7556175).
\end{acknowledgments}


\appendix

\section{Derivation of harmonic potential approximation}
\label{app:HarmonicPotentialApproximation}
As the Wigner crystal is in equilibrium, we can expand the potential $V(X_n,Y_n)$ around equilibrium positions of electrons. As a necessary condition for equilibrium, the first derivative of the potential is zero, so at lowest order we get a quadratic harmonic oscillator potential.\cite{meyer:jpcm09} We consider a quasi-1D Wigner crystal in the $XY$ plane with a metallic gate placed parallel to the $XY$ plane at a distance $d$ from the localized electrons. The gate models the effect of metallic gates usually present in experiments that screen the Coulomb interaction between electrons by generating ``image charges''. As a result of this screening of the long-range part of the Coulomb interaction, we consider only nearest neighbour Coulomb interactions between electrons. Our interaction potential then reads
\begin{widetext}
\begin{eqnarray}
\nonumber
&&V(x_n,y_n)=\frac{e^2}{\epsilon}\left[\frac{1}{2\sqrt{(a+x_{n,1}-x_{n-1,2})^2+(y_{n,1}-y_{n-1,2}-w)^2}}-\frac{1}{2\sqrt{(a+x_{n,1}-x_{n-1,2})^2+(y_{n,1}-y_{n-1,2}-w)^2+4d^2}}\right. \\ \nonumber&&\left.+\frac{1}{\sqrt{(a+x_{n,2}-x_{n,1})^2+(w+y_{n,2}-y_{n,1})^2}}-\frac{1}{\sqrt{(a+x_{n,2}-x_{n,1})^2+(w+y_{n,2}-y_{n,1})^2+4d^2}}\right]\\ &&+\frac{m\Omega_{con}^2}{2}\left[\left(y_{n,1}-\frac{w}{2}\right)^2+\left(y_{n,2}+\frac{w}{2}\right)^2\right].
\end{eqnarray}
\end{widetext}
The condition for equilibrium that the first derivative must be zero becomes
\begin{equation}
\label{eq:EquilibriumCondition}
\frac{e^2}{\epsilon}\left[\frac{2w}{(a^2+w^2)^{3/2}}-\frac{2w}{(a^2+w^2+4d^2)^{3/2}}\right]-\frac{m\Omega_{con}^2w}{2}=0.
\end{equation}
The potential after expansion has the form:
\begin{eqnarray}
\label{eq:VExpanded}
\nonumber&&V(x_n,y_n)\\ \nonumber&&\simeq[D_{xx}^{(1)}-D_{xx}^{(2)}][x_{n,1}-x_{n-1,2}]^2+[D_{xx}^{(1)}-D_{xx}^{(2)}][x_{n,2}-x_{n,1}]^2\\ \nonumber&&+[D_{yy}^{(1)}-D_{yy}^{(2)}][y_{n,1}-y_{n-1,2}]^2+[D_{yy}^{(1)}-D_{yy}^{(2)}][y_{n,2}-y_{n,1}]^2\\ \nonumber&&+[D_{xy}^{(2)}-D_{xy}^{(1)}][x_{n,1}-x_{n-1,2}][y_{n,1}-y_{n-1,2}]+\\ \nonumber&&+[D_{xy}^{(1)}-D_{xy}^{(2)}][x_{n,2}-x_{n,1}][y_{n,2}-y_{n,1}]+D_{\Omega}[y_{n,1}^2+y_{n,2}^2], \\
\end{eqnarray}
where
\begin{eqnarray}
\label{eq:Dxx1}
D_{xx}^{(1)}&=&\frac{e^2}{\epsilon}\frac{2a^2-w^2}{(a^2+w^2)^{5/2}},\\
D_{xx}^{(2)}&=&\frac{e^2}{\epsilon}\frac{2a^2-w^2-4d^2}{(a^2+w^2+4d^2)^{5/2}},\\
D_{yy}^{(1)}&=&\frac{e^2}{\epsilon}\frac{-a^2+2w^2}{(a^2+w^2)^{5/2}},\\
D_{yy}^{(2)}&=&\frac{e^2}{\epsilon}\frac{-a^2-4d^2+2w^2}{(a^2+4d^2+w^2)^{5/2}},\\
D_{xy}^{(1)}&=&\frac{e^2}{\epsilon}\frac{3aw}{(a^2+w^2)^{5/2}},\\
D_{xy}^{(2)}&=&\frac{e^2}{\epsilon}\frac{3aw}{(a^2+4d^2+w^2)^{5/2}},\\
D_{\Omega}&=&m\Omega_{con}^2.
\end{eqnarray}
In the following we consider the case where $d\gg a\gg w$, so that the most significant contributions are given by $D_{xx}^{(1)} \simeq 2e^2/(\epsilon a^3)$, and $D_{yy}^{(1)}\simeq -e^2/(\epsilon a^3)$. We also retain the confinement in the $Y$-direction, $D_{\Omega}$. The form of the potential we use is
\begin{equation}
\bar{V}(x_n,y_n)=\frac{m\Theta^2}{2}(x_n-x_{n+1})^2-\frac{m\Omega_1^2}{2}(y_n-y_{n+1})^2+\frac{m\Omega_2^2}{2}y_n^2.
\end{equation}

\section{Averaging out charge degrees of freedom for an arbitrary density of electrons}
\label{app:ArbitraryDensity}
Here we calculate $\langle S_{SOC}S_{SOC}\rangle$ for arbitrary $2m\alpha a$. From Eq.~(\ref{eq:HSOC}) we see that we must keep both the $\sigma_x^n\cos(2m\alpha a n)$ and $\sigma_z^n\sin(2m\alpha an)$ terms, so that going to Fourier space, the contribution to the action from SOC becomes
\begin{eqnarray}
\nonumber
S_{SOC}\simeq \frac{\alpha m}{2\beta N}\sum_{k,\omega} \omega y_{k,\omega}(\sigma_x^{2m\alpha-k,-\omega}+\sigma_x^{-2m\alpha-k,-\omega}\\ -i\sigma_z^{2m\alpha-k,-\omega}+i\sigma_z^{-2m\alpha-k,-\omega}).\ \ \
\end{eqnarray}
We perform the analogous calculation to that described in Section \ref{subsec:AveragingOutCharge} and keep only the nearest neighbor terms to obtain the result for $\langle S_{SOC}S_{SOC}\rangle$ that
\begin{eqnarray}
\label{eq:SOCSOCn}
\nonumber
\langle S_{SOC}S_{SOC}\rangle=\alpha^2m  \beta\frac{\Omega_1^2}{2\Omega_2}\sum_n\int_0^\beta d\tau\biggl[\sigma_x^{n,\tau}\sigma_x^{n+1,\tau}\\ \nonumber\times \left(\cos[2m\alpha a(2n+1)]+\cos[2m\alpha a]\right)\\ \nonumber+\sigma_x^{n,\tau}\sigma_z^{n+1,\tau}\left(\sin[2m\alpha a(2n+1)]+\sin[2m\alpha a]\right)\\ \nonumber+\sigma_z^{n,\tau}\sigma_x^{n+1,\tau}\left(\sin[2m\alpha a(2n+1)]-\sin[2m\alpha a] \right)\\ +\sigma_z^{n,\tau}\sigma_z^{n+1,\tau}\left(\cos[2m\alpha a]-\cos[2m\alpha a (2n+1)]\right)\biggr].\end{eqnarray}
Here we see that  for $2m\alpha a=(2l+1)\pi$ we recover to the previous result derived in Section \ref{subsec:AveragingOutCharge}. If we take $2m\alpha a=2l\pi$, the correction is positive, and consequently in the spin Hamiltonian the coefficient of $\sigma_x^n\sigma_x^{n+1}$ is smaller than the prefactor of $\sigma_y^n\sigma_y^{n+1}$ and $\sigma_z^n\sigma_z^{n+1}$. In this $XY$-phase, the XXZ model has a gapless spectrum.\cite{giamarchi:book}

In the case that $2m\alpha a=(2l+1)\pi/2$, the correcton arising from the SOC generates a nearest neighbor coupling between $\sigma_x$ and $\sigma_z$. To the best of our knowledge, the Hamiltonian of the form shown in Eq. (\ref{eq:SOCSOCn}) has not been studied, despite the fact that the investigation of complex similar models e.g. alternating Heisenberg chain,\cite{papenbrock:prb03} and models with Dzyaloshinskii-Moriya interactions \cite{affleck:prb99, oshikawa:prb02, gangadharaiah:prb08}  has been undertaken.

\section{Orbital parts of the wave functions for a double well potential}
\label{app:Wavefunctions}
We define the orbital parts of the wave functions as presented in Refs.~[\onlinecite{burkard:prb99, stepanenko:prb03, stepanenko:prb12, kornich:prb14}]. They are constructed from the ground state wave functions of the harmonic oscillator:
\begin{eqnarray}
\phi_{L,R}(x_n,y_n)=\frac{1}{\sqrt{\pi}l_c}\exp \left(-\frac{[(x_n \pm a/2)^2+y_n^2]}{2l_c^2} \right).
\end{eqnarray}
The wave functions for the electron in the left/right well are
\begin{eqnarray}
\Phi_{L,R}({\bm r}_n)=\frac{\phi_{L,R}(x_n,y_n)-g\phi_{R,L}(x_n,y_n)}{\sqrt{1-2sg+g^2}}\phi_Z(Z_n),\ \ \
\end{eqnarray}
where ${\bm r}_n=(x_n, y_n, Z_n)$, and $\phi_Z(Z_n)$ is a part of the wave function that depends on the confinement in the $Z$ direction, whose precise form is not important for us. We also define the quantities
\begin{eqnarray}
s&=&\langle\phi_L|\phi_R\rangle=\exp\left[-a^2/(4l_c^2)\right],\\
g&=&\frac{1-\sqrt{1-s^2}}{s}.
\end{eqnarray}
where $s$ gives the overlap between offset harmonic oscillator ground states centred on the left and right wells, and $g$ ensures orthogonality of the one-particle wave functions.

Constructing the two-particle wave functions from the one-particle ones we get
\begin{eqnarray}
\label{eq:Psipm}
\Psi_{\pm}({\bf r}_1,{\bf r}_2)&=&\frac{\Phi_L({\bf r}_1)\Phi_R({\bf r}_2)\pm\Phi_R({\bf r}_1)\Phi_L({\bf r}_2)}{\sqrt{2}},\\
\label{eq:PsiLR}
\Psi_{L,R}({\bf r}_1,{\bf r}_2)&=&\Phi_{L,R}({\bf r}_1)\Phi_{L,R}({\bf r}_2),
\end{eqnarray}
as used in the main body of the text.

\end{document}